\documentstyle{article}

 \input{amssym.def}
\input{amssym.tex} 
\newtheorem{Definition}{Definition}
\newtheorem{Lemma}{Lemma}
\newtheorem{Theorem}{Theorem}
\newtheorem{Proposition}{Proposition}

\renewcommand{\ll}{\label}
\newcommand{\be}{\begin{equation}}
\newcommand{\ee}{\end{equation}}
\newcommand{\bea}{\begin{eqnarray}}
\newcommand{\eea}{\end{eqnarray}}

\newcommand{\bib}{\bibitem}
\newcommand{\ci}{\cite}

\newcommand{\qm}{quantum mechanics}

\newcommand{\ca}{$C^*$-algebra}

\newcommand{\rep}{representation}
\newcommand{\irrep}{irreducible representation}
\newcommand{\Hs}{Hilbert space}

\newcommand{\ovl}{\overline}

\newcommand{\til}{\tilde}
\newcommand{\raw}{\rightarrow}

\newcommand{\n}{\parallel}

\newcommand{\la}{\langle}
\newcommand{\ra}{\rangle}

\renewcommand{\Im}{{\rm Im}\,}
\newcommand{\rst}{\upharpoonright}

\newcommand{\x}{\times}

\newcommand{\Ad}{{\rm Ad}}
\newcommand{\Co}{{\rm Co}}

\newcommand{\cin}{C^{\infty}}
\newcommand{\cci}{C^{\infty}_c}
\newcommand{\half}{\mbox{\footnotesize $\frac{1}{2}$}}

\newcommand{\Hh}{{\cal H}_{\hbar}}
\newcommand{\PHh}{{\Bbb P}{\cal H}_{\hbar}}
\newcommand{\PH}{{\Bbb P}{\cal H}}

\newcommand{\Ah}{{\frak A}^{\hbar}}

\newcommand{\q}{{\cal Q}_{\hbar}}
\newcommand{\lho}{\lim_{\hbar\rightarrow 0}}

\newcommand{\qh}{q_{\hbar}}

\newcommand{\qb}{{\cal Q}_{\hbar}^{B}}

\newcommand{\inv}{^{-1}}

\newcommand{\Exp}{{\rm Exp}}

\newcommand{\al}{\alpha}
\newcommand{\bt}{\beta}

\newcommand{\dl}{\delta}
\newcommand{\Dl}{\Delta}
\newcommand{\ep}{\epsilon}

\newcommand{\th}{\theta}

\newcommand{\lm}{\lambda}
\newcommand{\Lm}{\Lambda}
\newcommand{\rh}{\rho}
\newcommand{\sg}{\sigma}

\newcommand{\ta}{\tau}

\newcommand{\Ph}{\Phi}
\newcommand{\phv}{\varphi}

\newcommand{\ps}{\psi}
\newcommand{\Ps}{\Psi}
\newcommand{\om}{\omega}
\newcommand{\Om}{\Omega}
\newcommand{\A}{{\frak A}}
\newcommand{\B}{{\frak B}}
\newcommand{\GC}{{\frak C}}

\newcommand{\M}{{\frak M}}

\newcommand{\g}{{\frak g}}

\renewcommand{\t}{{\frak t}}
\renewcommand{\H}{{\cal H}}

\newcommand{\CN}{{\cal N}}

\newcommand{\CO}{{\cal O}}

\newcommand{\CQ}{{\cal Q}}

\newcommand{\CV}{{\cal V}}

\newcommand{\C}{{\Bbb C}}

\newcommand{\BP}{{\Bbb P}}

\newcommand{\N}{{\Bbb N}}
\newcommand{\R}{{\Bbb R}}

\makeatletter
\newskip\tempskip
\def\endproof{{\parfillskip24\p@ plus\@ne fil\@@par}\tempskip\prevdepth
  \ifdim\lastskip=\z@\tempskip\z@\else\vskip-\lastskip
    \ifdim\tempskip>4\p@ \tempskip.5\tempskip \else \tempskip\z@\fi\fi
  \nobreak\vskip-\baselineskip\vskip-\tempskip\noindent\hbox 
to\hsize{\hfill
    $\blacksquare$}\par\vskip\tempskip\vskip\abovedisplayskip\@doendpe}
\makeatother
\makeatletter
\newskip\tempskip
\def\endiproof{{\parfillskip24\p@ plus\@ne fil\@@par}\tempskip\prevdepth
  \ifdim\lastskip=\z@\tempskip\z@\else\vskip-\lastskip
    \ifdim\tempskip>4\p@ \tempskip.5\tempskip \else \tempskip\z@\fi\fi
  \nobreak\vskip-\baselineskip\vskip-\tempskip\noindent\hbox 
to\hsize{\hfill
    $\Box$}\par\vskip\tempskip\vskip\abovedisplayskip\@doendpe}
\makeatother
\newcommand{\enp}{\endproof}

\newcommand{\Ao}{\tilde{{\frak A}}_0} \renewcommand{\Ah}{{\frak
A}_{\hbar}} \topmargin = - 0.5 cm \textheight = 23 cm \textwidth = 15
cm \oddsidemargin = 0.9 cm
\begin{document} 
\setlength{\baselineskip}{1\baselineskip}
 \setlength{\unitlength}{1cm} \title{Strict
quantization of coadjoint orbits}
\author{N.P.~Landsman\thanks{Supported by a fellowship from the Royal
Netherlands Academy of Arts and Sciences (KNAW)}\\ \mbox{}\hfill \\
Korteweg-de Vries Institute for Mathematics \\ University of Amsterdam
\\ Plantage Muidergracht 24 \\ 1018 TV AMSTERDAM, THE NETHERLANDS \\
\mbox{}\hfill \\ {\em email:} npl@wins.uva.nl } \date{\today}
\maketitle
\begin{abstract}
A strict quantization of a compact symplectic manifold $S$ on a subset
$I\subseteq\R$, containing 0 as an accumulation point, is defined as a
continuous field of \ca s $\{\Ah\}_{\hbar\in I}$, with $\A_0=C_0(S)$,
and a set of continuous cross-sections $\{\CQ(f)\}_{f\in \cin(S)}$ for
which $\CQ_0(f)=f$.  Here $\q(f^*)=\q(f)^*$ for all $\hbar\in I$,
whereas for $\hbar\raw 0$ one requires that
$i[\q(f),\q(g)]/\hbar\raw\q(\{f,g\})$ in norm. We discuss general
conditions which guarantee that a (deformation) quantization in a more
physical sense leads to one in the above sense.

Using ideas of Berezin, Lieb, Simon, and others, we construct a strict
quantization of an arbitrary integral coadjoint orbit $\CO_{\lm}$ of a
compact connected Lie group $G$, associated to a highest weight
$\lm$. Here $I=0\cup 1/\N$, so that $\hbar=1/k$, $k\in\N$, and $\A_{1/k}$
is defined as the \ca\ of all matrices on the finite-dimensional \Hs\
$\CV_{k\lm}$ carrying the \irrep\ $U_{k\lm}(G)$ with highest weight
$k\lm$.  The quantization maps $\CQ_{1/k}(f)$ are constructed from coherent
states in $\CV_{k\lm}$, and have the special feature of being
 positive maps.  
\end{abstract}
1998 PACS: 02.20Qs, 03.65Db, 02.40Vh, 02.30Tb
\thispagestyle{empty}
\newpage
\section{Introduction}
The aim of this paper is to combine a number of what appear to the
author to be good ideas in mathematical physics, whose interplay has
so far not sufficiently been studied. 
Firstly, there now exists a satisfying $C^*$-algebraic definition of
quantization,  which enables one to link Poisson and symplectic geometry \ci{GS,MR}
with non-commutative geometry \ci{Con}. In particular, the geometric theory of
classical mechanics and reduction is thereby related to the
$C^*$-algebraic formulation of quantum mechanics and induction
\ci{MT}. The main mathematical idea of this definition goes back to Rieffel
\ci{Rie1}, who showed how deformation quantization makes sense in an operator-algebraic context.
The physical postulates, of course, may be traced back to Dirac.  
In this definition, like in conventional (`formal') deformation quantization, 
it is crucial that one studies the quantum theory for a family of values of Planck's constant
$\hbar$ \ci{Ber}.

Secondly, one has the well-known connection between coadjoint orbits
of certain Lie groups, unitary \rep s, and geometric quantization
\ci{Kos}. This connection works particularly well for either compact
or exponential nilpotent Lie groups. Here one keeps $\hbar$ fixed.

Our third source of inspiration is
 the work of Lieb \ci{Lie}, Simon \ci{Sim1}, and
others on the classical limit of quantum spin systems and their
generalizations to arbitrary compact Lie groups.
 This work is closely related to Perelomov's coherent states
\ci{Per1,Per2}, as well as to Berezin's approach to quantization and
the classical limit \ci{Ber}.

 Historical comments and
extensive references concerning these ideas  may be
found in \ci{MT}.

Using the second and the third group of ideas at both a conceptual and
a technical level, we will construct a $C^*$-algebraic quantization of
an arbitrary integral coadjoint orbit of a compact connected Lie
group. A different approach to this problem has recently been
considered in \ci{Haw}; also see \ci{Bar}.

Section 2 contains a general $C^*$-algebraic definition of
quantization. A quantization satisfying this definition is called `strict'.
 We present criteria on
a given quantization which guarantee that the postulates in the definition are met. It is on
this basis that we will actually construct the quantizations in this
paper. Section 3 describes our approach to coherent states and  Berezin quantization,
and develops conditions under which the  Berezin quantization constructed from
a family of coherent states is strict.
This is done in terms of a so-called pure state quantization, which, in a heuristic sense,
 is dual to a \ca ic quantization.
The material in sections 2 and 3 is model-independent, and should be relevant to quantization
theory in general.

Section 4 is a brief review of Perelomov coherent states,
coadjoint orbits of compact Lie groups, the momentum map, and the connection between 
these concepts. The material in this section is not new, but is worth summarizing in preparation for
our main results. Section 5 contains the two principal theorems of this paper.
 The first states that Perelomov's coherent states for compact connected  Lie groups
define pure state quantizations of particular coadjoint orbits. The second states that
such pure state quantizations lead to  strict quantizations of the coadjoint orbits in question.
\section{On quantization}
The central notion in $C^*$-algebraic quantization theory is that of a
continuous field of \ca s \ci{Dix}.  For our purposes the following
reformulation is useful \ci{KW}.
\begin{Definition}\ll{defcfca}
A continuous field of \ca s $(\GC,\{\A_x,\phv_x\}_{x\in X})$ over a
locally compact Hausdorff space $X$ consists of a \ca\ $\GC$, a
collection of \ca s $\{\A_x\}_{x\in X}$, and a set
$\{\phv_x:\GC\raw\A_x\}_{x\in X}$ of surjective
$\mbox{}^*$-homomorphisms, such that for all $A\in\GC$:
\begin{enumerate}
\item
the function $x\raw \n\phv_x(A)\n$ is in $C_0(X)$;
\item
one has $\n A\n=\sup_{x\in X}\n\phv_x(A)\n$;
\item
for any $f\in C_0(X)$ there is an element $fA\in\GC$ for which
$\phv_x(fA)=f(x)\phv_x(A)$ for all $x\in X$.
\end{enumerate}
\end{Definition}
Here $C_0(X)$ is the space of continuous functions on $X$ which vanish
at infinity.
Note that it is sufficient that each $\phv_x$ has dense image, since
the image of a $\mbox{}^*$-homomorphism between \ca s is automatically
closed.  The continuous cross-sections of the field in the sense of
\ci{Dix} consist of those elements $\{A_x\}_{x\in X}$ of $\prod_{x\in
X}\A_x$ for which there is a $A\in \GC$ such that $A_x=\phv_x(A)$ for
all $x\in X$.

The following definition of quantization, which is a slight
 reformulation of a definition in \ci{Rie5}, seems to combine the best
 of previous definitions in this direction in \ci{Rie1} and \ci{NPL1}.
\begin{Definition}\ll{defqua}
Let $I\subseteq\R$ contain $0$ as an accumulation point.  A strict
quantization of a Poisson manifold $P$ on $I$ consists of
\begin{enumerate}
\item 
a continuous field of \ca s $(\GC,\{\Ah,\phv_{\hbar}\}_{\hbar\in I})$ over $I$,
with $\A_0=C_0(P)$;
\item
a dense $\mbox{}^*$-subalgebra $\til{\A}_0$ of $C_0(P)$ on which the
Poisson bracket is defined, and which is closed under taking Poisson
brackets (so that $\Ao$ is a complex Poisson algebra);
\item
a linear map $\CQ:\Ao\raw\GC$ which (with
$\q(f)\equiv\phv_{\hbar}(\CQ(f))$) for all $f\in \til{\A}_0$ and
$\hbar\in I$ satisfies \bea \CQ_0(f)& =& f, \ll{q0f}\\ \q(f^*)& =&
\q(f)^*,\ll{real} \eea and for all $f,g\in \til{\A}_0$ satisfies
Dirac's condition \be \lho \n \frac{i}{\hbar}[\q(f),\q(g)]
-\q(\{f,g\})\n =0. \ll{direq} \ee
\end{enumerate}
\end{Definition}
Elements of $I$ are interpreted as possible values of Planck's
constant $\hbar$, and $\Ah$ is the quantum algebra of observables of
the theory at the given value of $\hbar\neq 0$. For real-valued $f$,
the operator $\q(f)$ is the quantum observable associated to the
classical observable $f$.  This interpretation is possible because of
condition (\ref{real}) in Definition \ref{defqua}.

In the examples of the present paper, the Poisson manifold $P$ will be
a compact symplectic manifold $S$, and we will choose
$\Ao=\cin(S)$. It will be a special feature that $\CQ$ is defined on
all of $\A_0$; the condition (\ref{direq}) of course makes sense on
$\Ao$ only.

The connection between Definition \ref{defqua} and the more physically
and historically oriented definition of quantization proposed in
\ci{NPL1,MT} is established in Proposition \ref{sdgcfca} below. This 
requires two lemmas,  the first of which, as a bonus,
 entails the equivalence between
the definition of a continuous field given above, and the one given by
Dixmier \ci{Dix} (restricted to the case that the base space is locally
compact).
\begin{Lemma}\ll{luclemma}
The \ca\ $\GC$ of (sections of) a continuous field is
 locally uniformly closed. That is,
 if $A\in \prod_x\A_x$ is such that for every
$y\in X$ and every $\ep>0$ there exists a $B^{y}\in\GC$ and a
neighbourhood $\CN^{y}$ of $y$ in which $\n A_x-B^{y}_x\n\,< \ep$ for all
$x\in \CN^y$, and also $\lim_{x\raw\infty} \n A_x\n=0$, then $A\in
\GC$.

 Alternatively, if the function $x\raw\n A_x-C_x\n$ lies
in $C_0(X)$ for each $C\in\GC$, then $A\in\GC$. 
\end{Lemma}

In the situation of the first part, there is a
compact set $K\subseteq X$ for which $\n A_x\n\,<\ep$ outside $K$,
as well as a
 finite subcover
$\{\CN^{x_1},\ldots \CN^{x_n}\}$ of $K$. Taking a  partition of unity
$\{u_i\}$ on $K$ subordinate to this subcover, the operator
$B=\sum_i u_i B^{x_i}$ lies in $\GC$ because of Definition
\ref{defcfca}.3, and satisfies
$\sup_{x\in X} \n A_x-B_x\n\,<\ep$. Hence $A\in\GC$ by Definition
\ref{defcfca}.2
and the completeness of $\GC$.

Given any $A\in\prod_x\A_x$ and $y\in X$, because $\phv_y$ is surjective
there is a $B^y\in\GC$ so that $A_{y}=B^y_y$. 
The assumption in the second part of the lemma then implies that the
conditions in the first part are satisfied, so that $A\in\GC$.
\enp

 The next lemma adapts Props.\ 10.2.3 and 10.3.2 in \ci{Dix}, which relate to
 Dixmier's own definition
of a continuous field,  to Definition \ref{defcfca}. 
\begin{Lemma}\ll{cfcalemma}
Suppose one has a family $\{\A_x\}_{x\in X}$ of \ca s indexed by a
locally compact Hausdorff space $X$, as well as a subset
$\til{\GC}\subseteq\prod_x\A_x$ which satisfies the following
conditions:
\begin{enumerate}
\item
 the set $\{A_x\, | \,A\in\til{\GC}\}$ is dense in $\A_x$ for each
 $x\in X$;
\item
the function $x\raw \n A_x\n$ is in $C_0(X)$ for each $A\in \til{\GC}$;
\item
 the set $\til{\GC}$ is closed under pointwise scalar multiplication,
 addition, adjointing, and operator multiplication.
\end{enumerate}

There exists a unique continuous field of \ca s
$(\GC,\{\A_x,\phv_x\}_{x\in X})$ whose collection of continuous
cross-sections contains $\til{\GC}$. Firstly, as a set $\GC$ consists
of all $A\in\prod_x\A_x$ for which the function $x\raw\n A_x-C_x\n$
lies in $C_0(X)$ for each $C\in\til{\GC}$.
 This set is regarded as a \ca\ under the pointwise
operations listed in item 3 above, and the norm defined in Definition
\ref{defcfca}.2. Secondly, $\phv_x(A)=A_x$ is the evaluation map.
\end{Lemma}
 
We first show that if $A\in \prod_x\A_x$ is such that for every
$x_0\in X$ and every $\ep>0$ there exists a $B\in\GC$ and a
neighbourhood $\CN$ of $x_0$ such that $\n A_x-B_x\n\,< \ep$ for all
$x\in \CN$, and also $\lim_{x\raw\infty} \n A_x\n=0$, then $A\in
\GC$. Indeed, take $C\in\til{\GC}$ arbitrary, and define the functions
$f_{AC}:x\raw \n A_x-C_x\n$ and $f_{BC}:x\raw \n B_x-C_x\n$.  Using
the inequality \be \ |(\n X\n-\n Y\n)|\leq\, \n X-Y\n, \ll{goodineq}
\ee one finds $|f_{AC}(x)-f_{BC}(x)|\,<\ep$ for all $x\in\CN$. By
assumption, $f_{BC}$ is continuous, so that
$|f_{BC}(x)-f_{BC}(x_0)|\,<\ep$ for all $x$ in some neighbourhood
$\CN'$ of $x_0$.  Combining the two inequalities yields
$|f_{AC}(x)-f_{AC}(x_0)|\,<3 \ep$ for all $x\in\CN\cap\CN'$.  Hence
$f_{AC}$ is continuous at $x_0$, which was arbitrary, so that
$A\in\GC$ by definition of $\GC$.
 
Using this property, it is easily shown that $\GC$ is a \ca, and that
 condition 3 in Definition \ref{defcfca} is satisfied. 
It is clear from  Definition \ref{defcfca}.1 and the definition 
of $\GC$ that
 $\GC$ is maximal. On the other hand,
according to the second part of Lemma \ref{luclemma}, $\GC$ is minimal,
so that it is unique. \enp

Adding an assumption satisfied by the examples in this paper,
we are now in a position to relate Definitions \ref{defcfca} and
\ref{defqua}.
\begin{Proposition}\ll{sdgcfca}
Suppose one has a Poisson manifold $P$ and a family $\{\Ah\}_{\hbar\in
I}$ of \ca s, with $\til{\A}_0\subset\A_0=C_0(P)$ as in Definition
\ref{defqua}, and a collection of linear maps $\{\q:\til{\A}_0\raw
\Ah\}_{\hbar\in I}$ satisfying (\ref{q0f}), (\ref{real}),
(\ref{direq}), as well as  
 \bea
& \lho & \n \half(\q(f)\q(g)+\q(g)\q(f))-\q(f g)\n =0; \ll{vneq} \\
& \lho &\n \q(f)\n =\n f\n_{\infty} \ll{bohreq}
\eea
 for all $f,g\in\Ao$, and finally the completeness condition that the
collection $\{\q(f)\, | \,f\in\Ao\}$ be dense in $\Ah$ for each
$\hbar\in I$. Furthermore, assume that $\hbar$ assumes discrete values.

 There exists a unique continuous field of \ca s
$(\GC,\{\Ah,\phv_{\hbar}\}_{\hbar\in I})$ whose collection of
continuous cross-sections $\{\phv_{\hbar}(A)\}_{\hbar\in I}$,
$A\in\GC$, contains all maps $\{\q(f)\}_{\hbar\in I}$, $f\in\Ao$.
\end{Proposition}

One defines $\til{\GC}\subset\prod_{\hbar}\Ah$ as the complex linear
span of all expressions of the form $\hbar\raw\q(f_1)\ldots\q(f_n)$,
where $f_i\in \Ao$.

Given that $I$ is discrete, the continuity of each function
$F_{f_1,\ldots,f_n}:\hbar\raw \n\q(f_1)\ldots\q(f_n)\n$ away from $0$
is trivial.  Now note that (\ref{vneq}) and (\ref{direq}) imply that
\be \lho \n \q(f) \q(g)-\q(f g)\n =0 \ll{cheq} \ee for all
$f,g\in\Ao$. Hence $\lho\n\q(f_1)\q(f_2\ldots f_n)-\q(f_1\ldots
f_n)\n=0$, and by induction $\lho\n\q(f_1)\ldots \q(f_n)-\q(f_1\ldots
f_n)\n=0$. Eq.\ (\ref{goodineq}) then yields $\lho \n\q(f_1)\ldots
\q(f_n)\n-\n\q(f_1\ldots f_n)\n=0$, so that finally $\lho
\n\q(f_1)\ldots \q(f_n)\n=\n f_1\ldots f_n\n_{\infty}$ by
(\ref{bohreq}).  Because of (\ref{q0f}), this shows that
$F_{f_1,\ldots,f_n}$ is continuous at $\hbar=0$.

Hence one is in the situation of Lemma \ref{cfcalemma}, and the claim
follows.  \enp

Although it is irrelevant for the present paper, we note that
Proposition \ref{sdgcfca} equally well holds (with a different proof)
if the set $I$ is not
discrete, provided that all $\Ah$ are identical for $\hbar\neq 0$, and
in addition the function $\hbar\raw\q(f)$ is continuous for each
$f\in\til{\A}_0$ \ci{MT}.
\section{Quantization of pure states and Berezin quantization} 
A \Hs\ $\H$ is a symplectic manifold, with symplectic form
$\om(\Ph,\Om)=2\Im (\Ph,\Om)$ (here the inner product $(\, ,\,)$ on
$\H$ is linear in the second entry) \ci{GS,MR}. This form is invariant
under the standard action $\exp(i\al):\Ps\raw\exp(i\al)\Ps$ of $U(1)$
on $\H$, so that the quotient $\H^*/U(1)$ is a Poisson manifold (here
$\H^*=\H\backslash\{0\}$).  The symplectic leaves \ci{MR} of
$\H^*/U(1)$ are the spaces $S_r=\H_r/U(1)$, where $\H_r=\{\Ps\in\H\,
|\, (\Ps,\Ps)=r^2\}$. In particular, the projective space $\PH$ may be
identified with $S_1$, and is therefore a symplectic manifold, with
symplectic form $\om_{\PH}$.

In addition, $\PH$ is equipped with a transition probability $p:\PH\x\PH\raw[0,1]$, given by
$p(\ps,\phv)=|(\Ps,\Ph)|^2$; here $\Ps$ and $\Ph$ are arbitrary lifts of $\ps$ and $\phv$ to
unit vectors in $\H$.
Equipped with these transition probabilities and with the Poisson bracket defined by 
 $\om_{\PH}$, the manifold $\PH$ is the pure state space of a quantum system without superselection
rules. See \ci{MT} and refs.\ therein.

The pure state space of a classical system is a symplectic manifold
$(S,\om_S)$, supporting the Liouville measure $\mu_L$. Such a
classical pure state space may be seen as carrying the `classical'
transition probability $p_0$, defined by $p_0(\rh,\sg)=\dl_{\rh\sg}$.
\begin{Definition}\ll{defcohst} Let $I\subseteq \R$ be as in
Definition \ref{defqua}, and put $I_0=I\backslash\{0\}$.  A pure state
quantization of a symplectic manifold $(S,\om)$ consists of a
collection of \Hs s $\{\H_{\hbar}\}_{\hbar\in I_0}$ and a collection
of smooth injections $\{\qh:S\raw \PHh\}_{\hbar\in I_0}$, for which
the following requirements are satisfied.  \begin{enumerate}
\item
There exists a positive function $c:I_0\raw \R\backslash \{0\}$ such that
for all $\hbar\in I_0$ and all $\ps\in\PHh$ one has
\be
 c(\hbar) \int_S  d\mu_L(\sg)\, p(\qh(\sg),\ps)=1. \ll{qhnorm}
\ee
  \item
For all fixed $f\in C_c(S)$ and  $\rh\in S$ the function 
$$ \hbar\raw \int_S  d\mu_L(\sg)\, p(\qh(\rh),\qh(\sg))f(\sg)$$ is continuous on  $I_0$
 and satisfies
\be
\lho c(\hbar) \int_S  d\mu_L(\sg)\, p(\qh(\rh),\qh(\sg)) f(\sg)=f(\rh). \ll{qh1}
\ee
\item
for each $\hbar\in I_0$, the map $\qh$ is a symplectomorphism, that is,
\be
\qh^*\om_{\PH}=\om_S.\ll{qhsympl}
\ee
\end{enumerate}\end{Definition}
 
Since $f\in C_c(S)$, the continuity requirement on  $I_0$ in this condition 
is equivalent to the continuity of  the function $\hbar\raw p(\qh(\rh),\qh(\sg))$
 for fixed $\rh$ and $\sg$. Using Urysohn's lemma, it is not difficult to
show \ci{MT} that (\ref{qh1})  and (\ref{qhnorm}) imply
\be
\lho p(\qh(\rh),\qh(\sg)) =\dl_{\rh\sg}. \ll{prsdlrs}
\ee
That is, in quantizing pure states the
quantum-mechanical transition probabilities should converge to the classical ones for $\hbar\raw 0$.
 
More generally, one may replace the Liouville measure $\mu_L$ in 
Definition \ref{defcohst} by a family of measures $\mu_{\hbar}$,
and substitute an appropriate limiting condition for (\ref{qhsympl});
see \ci{MT}. This generality is not needed for the present paper.
The author is indebted to E. Hawkins for this remark.

A  pure state quantization naturally leads to the quantization of observables.
\begin{Definition}\ll{defBT}
Let  $\{\H_{\hbar},\qh\}_{\hbar\in I_0}$ be  a pure state quantization of a symplectic
manifold $S$. The  Berezin quantization  of a function $f\in
L^{\infty}(S)$ is the family of operators $\{\qb(f)\}_{\hbar\in I_0}$, where
$\qb(f)\in\B(\Hh)$ is defined  by the weak integral
\be
\qb(f)= c(\hbar) \int_S  d\mu_L(\sg) f(\sg) [\qh(\sg)]. \ll{btop}
\ee
Here $[\qh(\sg)]$ is the projection onto the one-dimensional subspace in $\Hh$ whose image in
$\PHh$ is $\qh(\sg)$. In case that $f\in L^1(S)\cap L^{\infty}(S)$, the integral
is an ordinary Lebesgue integral (with values in a Banach space).
\end{Definition}

A number of properties of $\qb$ are immediately evident (cf.\ \ci{Sim1}): 
$\qb$ is positive (that is, $f\geq 0$ almost everywhere on $S$ implies $\qb(f)\geq 0$ in
$\B(\Hh)$), and if $f$ is real-valued then $\qb(f)$ is self-adjoint.
Moreover, $\qb(f)$ is bounded, with
\be
\n\qb(f)\n\,\leq\,\n f\n_{\infty}. \ll{qbnorm}
\ee
To prove the last property, let $A$ be a bounded symmetric 
operator such that $|(\Ps,A\Ps)|\leq c\n\Ps \n^2$ for some $c>0$, and
all $\Ps$.  One then replaces $\Ps$ by $\Ps\pm
A\Ps/c$, and subtracts the two inequalities thus obtained.  This
implies the inequality $\n A\Ps\n\,\leq\ c\n\Ps\n$, showing that $A$
is bounded with norm $\leq c$.  This argument with (\ref{qhnorm})
implies (\ref{qbnorm}). 

A Berezin quantization is not necessarily strict when restricted to, say, $\cci(S)$; even
(\ref{qhsympl}) does not imply (\ref{direq}). At the present abstract level, all that can be
inferred is the following. 
\begin{Proposition}\ll{qbisstrict} 
Let $f\in C_0(S)$, and assume that $I$ is discrete. Then  
\be
\lho  \n \qb(f)\n =\n f\n_{\infty}.\ll{bohreqforbt}
\ee
\end{Proposition}
 
We initially assume that $f\in L^1(S)\cap C_0(S)$, and at the
end extend the result to $f\in C_0(S)$ using the continuity of
$\qb$. Eq.\ (\ref{qbnorm}) implies
\be
\lim\,\sup_{\hbar\raw 0} \n \qb(f)\n\,\leq\, \n f\n_{\infty}.\ll{lowersem}
\ee  
On the other hand, for $f\in C_0(S)$ we can find $\rh\in S$ for which $ \n f\n_{\infty}=|f(\rh)|$.
Now for any unit vector $\Ps\in\Hh$, eq.\ (\ref{btop}) implies that
\be
 (\Ps,\qb(f)\Ps)= c(\hbar) \int_S  d\mu_L(\sg)\, p(\qh(\sg),\ps) f(\sg),\ll{defqbtgel}
\ee 
where $\ps$ is the projection of $\Ps$ to ${\Bbb P}\Hh$. We now 
use the  obvious inequality $\n \qb(f)\n\, \geq\, |(\Ps,\qb(f)\Ps)|$,
take $\Ps$ to be a lift of $\qh(\rh)$, and use (\ref{qh1}) to find
\be
 \lim\,\inf_{\hbar\raw 0}  \n\qb(f)\n\, \geq\, \n f\n_{\infty}.\ll{uppersem}
\ee
Combining
 this with (\ref{lowersem}), eq.\ (\ref{bohreqforbt}) follows. \enp 
\section{Coadjoint orbits and Perelomov's coherent states}
 Let  $G$ be a Lie group, with Lie algebra $\g$.
The dual $\g^*$ of $\g$ is a Poisson manifold under the so-called Lie-Poisson bracket \ci{MR,MT}
\be
\{f,g\}(\th)=- \th([df_{\th},dg_{\th}]); \ll{liepbr}
\ee
here the differential $df_{\th}$ of $f\in\cin(\g^*)$ at $\th\in\g^*$, which is a linear map
from $T_{\th}\g^*\simeq \g^*$ to $\R$, is identified with an element of $\g\simeq\g^{**}$, so that
the right-hand side of (\ref{liepbr}) is the Lie bracket in $\g$.

The symplectic leaves of $\g^*$ with respect to the Lie-Poisson structure are the coadjoint orbits
of $G$ \ci{MR,MT}. This endows each coadjoint orbit $\CO$ with the so-called Lie-Kirillov symplectic
structure. It is clear that  $\th\in\g^*$ satisfies $\th([X,Y])=0$ for all $X,Y\in \g_{\th}$,
where $\g_{\th}$ is the Lie
algebra  of the stabilizer $G_{\th}$ of $\th$ under the coadjoint action. In other words,
$\th: \g_{\th}\raw\R$ is a Lie algebra homomorphism.
\begin{Definition}\ll{defintorbit}
A coadjoint orbit $\CO\in\g^*$ is called integral 
if for some (hence all) $\th\in\CO$ the
functional $\th\rst \g_{\th}$ exponentiates to a character of $G_{\th}$.\end{Definition}

 In other words, $\th$ is integral iff there is a character $U_{\th}$ of 
$G_{\th}$ such that $\th=idU_{\th}$ on $\g_{\th}$. If this holds for one $\th\in\CO$ it holds for
all, since one has $U_{\Co(x)\th}=U_{\th}\circ \Ad(x\inv)$.
Here and in what follows, $\Co$ and $\Ad$ stand for the coadjoint and the adjoint action of $G$
on $\g^*$ and on $\g$, respectively.

In the remainder of this paper, $G$ is a compact connected Lie group.
We assume familiarity with the standard Cartan-Weyl description of the unitary \irrep s of $G$
\ci{Sim2}. We fix a maximal torus $T\subset G$. The coadjoint orbits of $T$ in the dual
$\t^*$ of the Lie algebra $\t$ of $T$ are points. The integral coadjoint orbits of $T$
form the  weight lattice $\Lm\subset\t^*$.
The Weyl group $W=N(T)/T$ (where $N(T)$ is the normalizer of $T$)
  acts on $T$ by conjugation. The derivative of this  action 
is a $W$-action on $\t$, whose dual action  on
$\t^*$ is the projection of the coadjoint action of  $N(T)$ to  $N(T)/T$. 
This action maps the weight lattice $\Lm$ into itself. 

A functional $\lm\in\t^*$ is  regular  when
$w\lm=\lm$ for $w\in W$ implies $w=e$ (and singular otherwise); this
defines the sets $\t^*_{\mbox{\tiny r}}$ and $\Lm_{\mbox{\tiny
r}}=\t^*_{\mbox{\tiny r}} \cap
\Lm$ of regular elements and  regular weights in $\t^*$, respectively. 
A Weyl chamber is a connected component $C$ of $\t^*_{\mbox{\tiny r}}$, and thereby    
 forms an open convex cone in $\t^*$. Singular weights clearly lie
 on the boundary of some Weyl chamber.
 One  singles out  an arbitrary Weyl chamber $C_{\mbox{\tiny d}}$, and declares a weight
 dominant if it lies in the closure $\ovl{C}_{\mbox{\tiny d}}$. The point is now that each 
$W$-orbit intersects a given closed Weyl chamber  $\ovl{C}$ in exactly one point. 

Combining the Cartan-Weyl theory with that of Kostant \ci{Kos}, one obtains a number of
parametrizations of the unitary dual of $G$ (i.e., the set of unitary \irrep s of $G$ modulo
unitary equivalence).
 \begin{Proposition}\ll{kostant} 
 There exist bijective correspondences between
the unitary dual $\hat{G}$ of $G$, the set of $W$-orbits in $\Lm$,
the set $\Lm_{\mbox{\tiny d}}=\Lm\cap \ovl{C}_{\mbox{\tiny d}}$ of dominant weights,
and  the set of integral coadjoint orbits in $\g^*$. The latter set is isomorphic to $\t^*/W$.
In other words,  one has
\be
\hat{G}\simeq \Lm/W \simeq \Lm_{\mbox{\tiny d}}\simeq 
(\g^*/W)_{\mbox{\tiny integral}}.
\ee
\end{Proposition}

Though exceedingly  well-known, we recall the explicit form of the bijection $\hat{G}\simeq
\Lm_{\mbox{\tiny d}}$ in Proposition \ref{kostant}, as it will play an important role in the proof
of our main result. We use the standard Cartan-Weyl basis 
$\{H_j\}_{j=1,\ldots,r}\cup \{E_{\al},E_{-\al}\}_{\al\in\Dl^+}$ of $\g_{\C}$, where $r$ is the rank
of $G$, and  $\Dl^+$ is the set of positive roots (relative
to a choice of $C_{\mbox{\tiny d}}$). The vectors $H_i$ lie in $\t$, and the $E_{\pm\al}$ are
eigenvectors of each $H_i$ under restriction from $\g$ to $\t$ of the adjoint \rep\ of $\g$ on
$\g_{\C}$.

 A \Hs\ $\CV_{\lm}$ carrying an \irrep\ $U_{\lm}(G)$
corresponding to a dominant  weight $\lm\in\Lm_{\mbox{\tiny d}}$ has a 
highest weight vector $\Ps_{\lm}$ of norm 1, unique up to a phase, such that 
\be
dU_{\lm}(H_j)\Ps_{\lm}=-i\lm(H_j)\Ps_{\lm} \ll{hwvector1}
\ee
 for $j=1,\ldots,r$, whereas for all $\al\in\Dl^+$ one has
\be
dU_{\lm}(E_{\al})\Ps_{\lm}=0 .\ll{hwvector2}
\ee
 Since $dU_{\lm}(E_{\al})^*=-dU_{\lm}(E_{-\al})$, it follows that
\be
(\Ps_{\lm},dU_{\lm}(E_{\al})\Ps_{\lm})=0 \ll{hwvector3}
\ee
for all $\al\in\Dl$.
 
The association of a coadjoint orbit in $\g^*$ with a $W$-orbit in $\t^*$ is as follows.
 We write $\t_{\C}^{\perp}$ for the span of all $E_{\pm\al}$, and 
$\t^{\perp}=\t_{\C}^{\perp}\cap\g$. Given a weight $\lm\in\t^*$, 
putting $\th(\lm)=0$ on $\t^{\perp}$ and $\th(\lm)=\lm$ on
$\t$, one obtains an extension $\th(\lm)\in\g^*$ of $\lm$.   Thus the coadjoint orbit
$\CO_{\lm}=\CO_{\th(\lm)}$  associated to $\lm$ is  the coadjoint orbit through $\th(\lm)$; it is
obvious from the definition of the $W$-action on $\t^*$ that all points of $W$-orbit of $\lm$ are
mapped into $\CO_{\lm}$.  We will use the label $\lm$ to denote a
dominant weight in $\Lm_{\mbox{\tiny d}}\subset \t^*$, as well as the
corresponding element $\th(\lm)$ in the coadjoint orbit
$\CO_{\lm}\subset\g^*$.  

The correspondence $\Lm_{\mbox{\tiny d}}\simeq \g^*/G$ is most easily
described in terms of the momentum map \ci{GS,MR}.  The \rep\
$U_{\lm}(G)$ on $\CV_{\lm}$ quotients to a $G$-action $\til{U}_{\lm}$
on the projective space ${\Bbb P}\CV_{\lm}$.  Since $U_{\lm}$ is unitary,
this quotient action may be computed from the action of 
$U_{\lm}(G)$ on unit vectors in $\CV_{\lm}$.
As explained at the
beginning of section 3, ${\Bbb P}\CV_{\lm}$ is a symplectic
manifold. It is clear from the definition of the symplectic structure
on ${\Bbb P}\CV_{\lm}$ and the unitarity of $U_{\lm}$ that the reduced
$G$-action $\til{U}_{\lm}$ preserves the symplectic form $\om_{{\Bbb
P}\CV_{\lm}}$. More strongly, this action admits an equivariant
momentum map $J:{\Bbb P}\CV_{\lm}\raw \g^*$, given by \be
J_X(\ps)=i(\Ps,dU_{\lm}(X)\Ps), \ll{mmhwetc} \ee As always, the unit
vector $\Ps\in\CV_{\lm}$ is an arbitrary lift of $\ps\in {\Bbb
P}\CV_{\lm}$.
\begin{Proposition}\ll{mmcartan}
The coadjoint orbit $\CO_{\lm}$ corresponding to 
 an \irrep\ $U_{\lm}$ with highest weight vector $\Ps_{\lm}$ 
contains $J\ps_{\lm}$.   
In fact, the restriction of $J$ to ${\Bbb P} U_{\lm}(G)\Ps_{\lm}$  is a 
symplectomorphism onto $\CO_{\lm}$. \end{Proposition}

Here it is understood that  ${\Bbb P} U_{\lm}(G)\Ps_{\lm}$
inherits the  usual symplectic structure of ${\Bbb P}\CV_{\lm}$, and that  $\CO_{\lm}$ is
endowed with the  Lie-Kirillov symplectic form. For convenience, as well as for later reference,
 we include a proof of this fundamental result (also cf.\ \ci{GS} or \ci{MT}).

Eq.\  (\ref{mmhwetc}),  (\ref{hwvector1}), (\ref{hwvector2}), and (\ref{hwvector3}) imply that
$\la J(\ps_{\lm}),X\ra$ equals $\lm(X)$ for $X\in\t$ and equals 0 for $X\in\t^{\perp}$. 
 Hence  $J(\ps_{\lm})$ is precisely  the element $\th(\lm)\in\g^*$
discussed after (\ref{hwvector3}), proving the first claim.  

By (\ref{mmhwetc}), the stability group $G_{J(\ps_{\lm})}$ of $J(\ps_{\lm})$ consists
of those $x\in G$ for which 
\be
(U_{\lm}(x)\Ps_{\lm},dU_{\lm}(Y)U_{\lm}(x)\Ps_{\lm})=
(\Ps_{\lm},dU_{\lm}(Y)\Ps_{\lm})
\ee
 for all $Y\in\g$. 
Since $U_{\lm}$ is irreducible this implies that $\Ps_{\lm}$ and $U_{\lm}(x)\Ps_{\lm}$ define the
same element of $\PH_{\lm}$, proving that $G_{J(\ps_{\lm})}\subseteq G_{\ps_{\lm}}$.
The opposite inclusion is trivial from the equivariance of $J$. 
\enp 

The unit vectors in $\CV_{\lm}$ of the form $U_{\lm}(x)\Ps_{\lm}$, where $x\in G$, are Perelomov's
coherent states \ci{Per1,Per2}, which are parametrized by $G$. 
It should be clear by now that we are not interested in these
states themselves, but in their projections  to ${\Bbb P}\CV_{\lm}$.
It follows from Proposition \ref{mmcartan} that the ensuing family is parametrized by
the coadjoint orbit $\CO_{\lm}$. Remarkably, the  coadjoint orbits  
 of maximal dimension among all coadjoint orbits $\g$, which are the ones for which
 $G_{\lm}=T$, are precisely the ones labeled by a regular weight $\lm$. 
The ones of smaller dimension, for which $G_{\lm}$ properly contains $T$, are labeled
by a singular weight.
 \section{Berezin quantization of coadjoint orbits\ll{bqcor}}
Proposition \ref{mmcartan} shows how to construct the coadjoint orbit $\CO_{\lm}$ corresponding to a
given unitary \irrep\ $U_{\lm}$ by Proposition \ref{kostant}. The converse passage is accomplished
by the Borel-Weil-Bott theory, which may be seen as a special case of geometric quantization
\ci{Kos}. In what follows we assume the correspondence in question to be given, and instead are
interested in constructing an entire family of \rep s of $G$ that are in some sense associated to a
given coadjoint orbit. The fundamental idea of rescaling the label of an \irrep\ by multiplying
with $1/\hbar$ (which accordingly has to be quantized in the compact case) is due to Berezin
[1975a];
 also cf.\  Perelomov [1986]. In a more intuitive setting, this rescaling was explicit in the early
years of \qm, and seems to comprise one of the faces of Bohr's correspondence principle.
The idea was further developed in the context of the classical limit of \qm\ by Lieb \ci{Lie} and
Simon \ci{Sim1}.
\begin{Theorem}\ll{bqcoadthm} Let $G$ be a
compact connected Lie group, and $\CO_{\lm}$ an integral coadjoint
orbit (cf.\
\ref{defintorbit}), corresponding to a highest weight $\lm\in\Lm_{\mbox{\tiny d}}$.
For $\hbar=1/k$, $k\in\Bbb N$, define $\H_{\hbar}=\CV_{\lm/\hbar}$, i.e., the carrier space of the
\irrep\ $U_{\lm/\hbar}(G)$ with highest weight $\lm/\hbar=k\lm$.

Let $\ta:\Hh\raw\BP\Hh$ be the canonical projection.
The map $\qh:\CO_{\lm}\raw \BP\Hh$, given by 
\be
\qh(\Co(x)\lm)=\ta(U_{\lm/\hbar}(x)\Ps_{\lm/\hbar}), \ll{defqhcox}
\ee
is well defined and injective, and provides a pure state quantization of $\CO_{\lm}$ 
(equipped with minus the Lie-Kirillov symplectic structure) on $I=1/\Bbb N\cup 0$. 
\end{Theorem}

One should note here that $k\lm\in \ovl{C}_{\mbox{\tiny d}}$ when
$\lm\in \ovl{C}_{\mbox{\tiny d}}$, since Weyl chambers are convex
cones.  The map $\qh$ is well-defined and injective by the equation
$G_{k\lm}=G_{\lm}$ plus the argument on stability groups used in the
proof of Proposition \ref{mmcartan}. In fact, if we define
$J_{\hbar}:\BP\Hh\raw \g^*$ by (\ref{mmhwetc}) with the right-hand
side divided by $k$, it follows from \ref{mmcartan} that $J_{\hbar}$
takes values in $\CO_{\lm}$ and is a left-inverse of $\qh$.

We start from  the fact that
the Haar measure on $G$ (with total mass 1) pushes forward to the Liouville measure derived from the
Lie symplectic structure under the canonical projection  $ G\raw\CO_{\lm}\simeq G/G_{\lm}$.
Using the invariance of the Haar measure and the unitarity of $U_{k\lm}$, 
we then have 
\be
\int_{\CO_{\lm}} d\mu_L(\sg)\, p(\qh(\rh),\qh(\sg)) f(\sg)=\int_G dx\, 
|(\Ps_{k\lm},U_{k\lm}(x)\Ps_{k\lm})|^2  f_{\lm}(yx) \ll{duffieldeq}
\ee
 for all $f\in C(\CO_{\lm})$,
where $f_{\lm}=\ta^*_{G\raw G/G_{\lm}}f\in C(G)$ is a right-$G_{\lm}$-invariant function, and   
$y$ is such that $\ta_{G\raw G/G_{\lm}}(y)=\rh\in\CO_{\lm}$. Choosing $f=1$, the orthogonality
relations  for compact groups then imply (\ref{qhnorm}) with  
\be
c(\hbar)=d_{\lm/\hbar},
\ee
 i.e., the dimension of $\Hh=\CV_{\lm/\hbar}$. 

Eq.\ (\ref{qhsympl}) follows from Proposition \ref{mmcartan}.
To prove (\ref{qh1}) we use a result of Gilmore \ci{Gil}.
\begin{Lemma}\ll{gilmore}
Let   $\lm_i$ be dominant weights with highest weight \rep s and vectors
$U_{\lm_i}$ and $\Ps_{\lm_i}$, respectively ($i=1,2$). Then for each $x\in G$
\be
(\Ps_{\lm_1},U_{\lm_1}(x)\Ps_{\lm_1})\cdot (\Ps_{\lm_2},U_{\lm_2}(x)\Ps_{\lm_2})=
(\Ps_{\lm_1+\lm_2},U_{\lm_1+\lm_2}(x)\Ps_{\lm_1+\lm_2}).
\ee \end{Lemma}

This is immediate from (\ref{hwvector1}), (\ref{hwvector2}), and the connectedness of $G$.\enp 

This lemma implies that 
\be
(\Ps_{k\lm},U_{k\lm}(x)\Ps_{k\lm})=(\Ps_{\lm},U_{\lm}(x)\Ps_{\lm})^k.\ll{gilmoreeq}
\ee
Using (\ref{duffieldeq}), we can write the
left-hand side of (\ref{qh1})   as 
$$
\lho\int_{\CO_{\lm}} d\mu_L(\sg)\, p(\rh_{\hbar},\sg_{\hbar}) f(\sg)=\lim_{k\raw\infty}  \int_{G}
d\mu_k(x)f_{\lm}(yx),
$$
 where  $\mu_k$ is a  probability measure on $G$ defined by 
\be
d\mu_k(x)= d_{k\lm}\, dx |(\Ps_{\lm},U_{\lm}(x)\Ps_{\lm})|^{2k}.
\ee
It is obvious that each $\mu_k$ is right-$G_{\lm}$-invariant. 
It follows from  (\ref{hwvector2}), (\ref{hwvector3}), and the fact that the exponential map is
surjective for compact   Lie groups, that $|(\Ps_{\lm},U_{\lm}(x)\Ps_{\lm})|$, which is
evidently $\leq 1$,  equals 1 iff $x\in G_{\lm}$. Hence for large $k$ the support of $\mu_k$ is
increasingly concentrated on $G_{\lm}$. This suggests that
\be
\lim_{k\raw\infty}\mu_k(f)=\int_{G_{\lm}}dh\, f(h) \ll{duffield}
\ee
 for all $f\in C(G)$, where $dh$ is
the normalized Haar measure on $G_{\lm}$. This is indeed the case, as proved by Duffield
\ci{Duf} (related results may be found in \ci{Lie} and \ci{Sim1}).
 For the right-$G_{\lm}$-invariant function $f_{\lm}\in C(G)^{G_{\lm}}$
one therefore obtains  $$\lim_{k\raw\infty}\mu_k(f_{\lm})=f_{\lm}(e).$$  

This proves (\ref{qh1}), which finishes the proof of Theorem
\ref{bqcoadthm}.\enp 

The Berezin quantization $\qb$ associated with the pure state quantization in
Theorem \ref{bqcoadthm} is defined on $\til{\A}_0=\cin(\CO_{\lm})$. By (\ref{btop}), one has
\be
\CQ^B_{1/k}(f)= d_{k\lm} \int_G dx\, f_{\lm}(x)[U_{k\lm}(x)\Ps_{k\lm}]; \ll{QBgroup}
\ee
this is an element of $\A_{1/k}=\M_{d_{k\lm}}(\C)$ (the \ca\ of complex $d_{k\lm}\x d_{k\lm}$
matrices). Recall that $[\Phi]$ is the projection onto $\C\Ph$. 
 
The most important property of $\qb$ is its $G$-equivariance. For $x\in G$ we
write  \be
\al_x^{1/k}(A)=U_{k\lm}(x)AU_{k\lm}(x)^*,\ll{alx1n}
\ee
where $A\in \A_{1/k}$, and for $f\in\A_0=C(\CO_{\lm})$ we put
\be
\al_x^0(f)=\Co(x\inv)^* f. \ll{alxclas}
\ee
\begin{Proposition}\ll{eqqbgroup}
For all $k\in \Bbb N$, $x\in G$, and $f\in L^{\infty}(\CO_{\lm})$ one has 
\be
\CQ^B_{1/k}(\al_x^0(f))=\al_x^{1/k}(\CQ^B_{1/k}(f)).\ll{eqqbgroupeq}
\ee \end{Proposition}

This is immediate from (\ref{QBgroup}),  (\ref{alxclas}), the fact that
$(\Co(y)^* f)_{\lm}=L_y^*f_{\lm}$,
 the right-invariance of the Haar measure, and (\ref{alx1n}). \enp

We are now ready for the main result of this paper.
\begin{Theorem}\ll{sqofcoad}
The Berezin quantization (\ref{QBgroup}), defined on the space
$\cin(\CO_{\lm})$, is strict. \end{Theorem}

We show that the assumptions in Proposition \ref{sdgcfca} are met.  
Eq.\ (\ref{bohreq})  follows from Theorem \ref{bqcoadthm} and Proposition
\ref{qbisstrict}. The completeness condition   is an easy consequence of
Schur's lemma and the irreducibility of $U_{k\lm}$.

We will now  prove  the remaining conditions (\ref{vneq}) and (\ref{direq}),
using the notation of the proof of Theorem \ref{bqcoadthm}.
 We pick a unit vector
$\Ph_k$ in each $\CV_{k\lm}$, and use the invariance of the Haar measure and 
the orthogonality relations for compact groups to write
\be
 \left(\Ph_k,(\CQ^B_{1/k}(f)\CQ^B_{1/k}(g)-\CQ^B_{1/k}(fg))\Ph_k\right) =  
 d_{k\lm}\int_G dx\,
f_{\lm}(x) (\Ph_k,U_{k\lm}(x)\Ps_{k\lm}) I_k(x),  \ll{qb1n}
\ee
where 
\be 
I_k(x)= d_{k\lm} \int_G dy\, (\Ps_{k\lm},U_{k\lm}(y)\Ps_{k\lm}) F^x_{\lm}(y),\ll{yintegration}
\ee
in which we have abbreviated
\be
 F^x_{\lm}(y)= (U_{k\lm}(xy)\Ps_{k\lm},\Ph_k)[g_{\lm}(xy)-g_{\lm}(x)].\ll{FgmyPh}
\ee

In the notation used after (\ref{duffieldeq}), the function $F^x_{\lm}$ on $G$ corresponds to a
function $F^x$ on $\CO_{\lm}$.

 Using (\ref{gilmoreeq}), we can write 
$(\Ps_{k\lm},U_{k\lm}(y)\Ps_{k\lm})=\exp(-kS_{\lm}(y))$, where $S_{\lm}(y)=-\log
(\Ps_{\lm},U_{\lm}(y)\Ps_{\lm})$ (in view of the exponentiation, the choice of the branch
cut of the logarithm is irrelevant). 
 The function $S_{\lm}$ is right-$G_{\lm}$-invariant; we denote
the corresponding function on $G/G_{\lm}$ by $S$.
We identify $G/G_{\lm}$ with $\CO_{\lm}$, so that the coset $[G_{\lm}]\in G/G_{\lm}$ is identified
with $\lm\in  \CO_{\lm}$.

 Putting $S_{\lm}^+(y)=-\log
|(\Ps_{\lm},U_{\lm}(y)\Ps_{\lm})|$, the absolute value of $\exp(-kS)$ is $\exp(-kS^+)$.
 As in the argument preceding (\ref{duffield}) we see that 
  $S^+$ takes values in $[0,\infty]$ and assumes its  unique absolute minimum $0$ at $\lm$. Since
$F^x_{\lm}$ in (\ref{yintegration}) is bounded, a standard argument implies that to $O(\exp(-k))$ we may
replace the integration over $G/G_{\lm}$ by one  over any neighbourhood of $\lm$.

We identify $T_{\lm}\CO_{\lm}$ with
$\g/\g_{\lm}$, and use complex co-ordinates $\{z_{\al},\ovl{z}_{\al}\}_{\al\in\Dl^+_{\lm}}$, where 
$\Dl^+_{\lm}$ consists of those positive roots for which $(\lm,\al)\neq 0$. By the definition of a
highest weight, this implies that $(\lm,\al)>0$ for all $\al\in\Dl^+_{\lm}$.
The co-ordinates $(z_{\al},\ovl{z}_{\al})$ correspond to the point in $\CO_{\lm}$ given by
$$ \Co\left[\Exp
\left(i\sum_{\al\in \Dl^+_{\lm}}(z_{\al}E_{\al}-\ovl{z}_{\al}E_{-\al})\right)\right]\lm.
$$ 

A simple computation, using the Cartan-Weyl form of the commutation relations
in $\g$ \ci{Sim2} and (\ref{hwvector1}) -
(\ref{hwvector3}), leads to
\be
S(z_{\al}, \ovl{z}_{\al})=\sum_{\al\in \Dl^+_{\lm}} (\lm,\al)z_{\al}
\ovl{z}_{\al}+O(|z|^3). 
\ee
Hence to $O(\exp(-k))$ we may approximate $I_k(x)$ by
$$ 
d_{k\lm} \int_{\g/\g_{\lm}} 
\left(
\prod_{\al\in \Dl^+_{\lm}} \frac{dz_{\al} d\ovl{z}_{\al}}{2\pi}
\right)\,
 J(z_{\al}, \ovl{z}_{\al}) e^{-k\sum_{\al\in
\Dl^+_{\lm}} (\lm,\al)z_{\al} \ovl{z}_{\al}}F^x(z_{\al}, \ovl{z}_{\al}),  
$$
where  $J$ is a Jacobian, and $F^x_{\lm}$ has been extended to $\g/\g_{\lm}$
by, say, the exponential map. If we omit the factor $[\ldots]$ in (\ref{FgmyPh}),
 the integral
(\ref{yintegration}) can be done, using the orthogonality relations
for compact groups. On the other
hand, we can compute the above integral to lowest order in the steepest descent approximation; this
avoids the need to compute $J(0)$. Comparing the results computes the prefactor in the  steepest
descent approximation as unity. As a by-product we obtain  the asymptotic expression for
$k\raw\infty$  
\be
d_{k\lm}\sim \frac{\prod_{\al\in \Dl^+_{\lm}} (\lm,\al)}{J(0)} k^{\half\dim(\CO_{\lm})},
\ll{weylas} 
\ee
where $\dim(\CO_{\lm})$ is given by 
\be
\dim(\CO_{\lm})=\dim(\g)-\dim(\t)-2\, {\rm Card}\, \{\al\in\Dl^+|(\lm,\al)=0\}.\ll{dimOgm}
\ee
 Comparison with the Weyl dimension formula \ci{Sim2}
then yields $J(0)=\prod_{\al\in \Dl^+}(\al,\dl)$, where $\dl=\half\sum_{\al\in \Dl^+}\al$.
Eq.\ (\ref{dimOgm}) follows from the decomposition
\be
\g_{\lm}=\t\bigoplus \oplus_{\al\in\Dl^+|(\lm,\al)=0}\g^{\R}_{\al}, \ll{chevalley}
\ee
where $\g^{\R}_{\al}=\g\cap(\C E_{\al} \oplus \C E_{-\al})$. This
decomposition is easily derived from the proof of Proposition
\ref{mmcartan}, the Cartan-Weyl commutation relations, and 
the  fact that $\lm(E_{\al})=0$ for all $\al\in\Dl$. 

Thus the  steepest descent approximation to  the above integral, and therefore to
(\ref{yintegration}), reads 
\be 
I_k(x)=\sum_{l=0}^N \frac{1}{l!} D^l(J F^x)(0) +O(k^{-N-1}), \ll{Inxis}
\ee
where, abbreviating $\partial_{\al}=\partial/\partial z_{\al}$ and $\ovl{\partial}_{\al}=\ovl{\partial}/\ovl{\partial} z_{\al}$,
we have put
\be 
D=\sum_{\al\in \Dl^+_{\lm}} \frac{1}{(\lm,\al)} \partial_{\al}\ovl{\partial}_{\al} . \ll{defDW}
\ee
Substituting this expansion in (\ref{qb1n}) we see that
\be
\left(\Ph_k,(\CQ^B_{1/k}(f)\CQ^B_{1/k}(g)-\CQ^B_{1/k}(fg))\Ph_k\right)=O(1/k). \ll{vnqb1n}
\ee
To analyze the remainder of $O(1/k)$ we note that the $l$'th term in the expansion leads to an
$x$-integrand in (\ref{qb1n})  of the form 
$$
(U_{k\lm}(x\inv)\Ph_k,\Ps_{k\lm})\ovl{(U_{k\lm}(x\inv)\Ph_k,\Ps^{(l_1)})} f \partial^{l_2}
\ovl{\partial}^{l_3} g_{\lm}(x),$$
where $l_i\leq l$ and the $\Ps^{(l_1)}$ are given by the action of products of $dU_{k\lm}(E_{\al})$
and  $dU_{k\lm}(E_{-\al})$ on $\Ps_{k\lm}$. The important point is now that
the orthogonality relations  (applied to the $x$-integration) then imply that the
$O(k^{-N-1})$ term is bounded by  $C \n\Ph_k\n^2/k^{N+1}$ for some constant $C$.   
Hence (\ref{vneq}) follows. 

To prove (\ref{direq}) we need the $l=1$ term in (\ref{Inxis}). We substitute 
(\ref{defDW}), and perform some partial integrations in the remaining $x$-integral (using the
invariance of the Haar measure). We abbreviate $A=(\Ph_k,U_{k\lm}(x)\Ps_{k\lm})$; then
(\ref{hwvector2}) implies that $\partial_{\al}A$ and $\ovl{\partial}_{\al}\ovl{A}$ vanish at
$z_{\al}=\ovl{z}_{\al}=0$.
 Terms of the form $\partial_{\al} \ovl{\partial}_{\al} A$ (or
$\ovl{A}$) drop out in the commutator, as do contributions from $J$ (whose first derivatives at 0
already vanish identically). What remains is
\be
\left(\Ph_k,ik [\CQ^B_{1/k}(f),\CQ^B_{1/k}(g)]-\CQ^B_{1/k}(\{f,g\})\Ph_k\right)=   O(1/k),
\ll{ditqb1n}
\ee
where, in the realization of $f,g$ as $G_{\lm}$-invariant functions $f_{\lm},g_{\lm}$ on $G$,
\be
\{f_{\lm},g_{\lm}\}=- i\sum_{\al\in \Dl_{\lm}}\frac{1}{(\lm,\al)}
 \xi^L_{\al} f_{\lm}\, \xi^L_{-\al}g_{\lm}. \ll{leipbeal}
\ee
Here the left-invariant vector fields $\xi^L_{\pm\al}$ on $G$ are defined by
first expressing $E_{\pm\al}\in\g_{\C}$ in terms of elements of $\g$, and then using the usual 
definition $\xi^L_Xf(y) =  df(y\Exp(tX)/dt(t=0)$.
  Also, $\Dl_{\lm}=\Dl_{\lm}^+\cup  \Dl_{\lm}^-$, i.e., the set of all roots $\al$ for which
$(\lm,\al)\neq 0$.

To finish the proof, we remark that (\ref{leipbeal}) is precisely the Lie-Kirillov Poisson bracket
on $\CO_{\lm}$; this may be verified at the point $\lm\in\CO_{\lm}$ (or $e\in G$) by direct
computation from (\ref{liepbr}), from which the general statement follows by the $G$-invariance of
the Poisson structure.

It is  manifest that the right-hand side of (\ref{leipbeal}) is left-$G$-invariant 
if $f_{\lm}$ and $g_{\lm}$ are. To prove the 
right-$G_{\lm}$-invariance of (\ref{leipbeal}), it is sufficient to establish
that it is right-invariant under the derived action of the Lie algebra $\g_{\lm}$, for
 $G_{\lm}$ is connected. Recall that
$[E_{\al},E_{\bt}]  =  N_{\al,\bt}E_{\al+\bt}$, where $\bt\neq -\al$. 
We now need the identity
 $N_{-\al-\bt,\bt}=- N_{\al,\bt}$ (where $\bt\neq\pm\al$), which follows from the
$\Ad(\g)$-invariance of the inner product on $\g_{\C}$, combined with the normalization of the
$E_{\al}$. The right-invariance  of (\ref{leipbeal}) under $\g_{\lm}$ follows by combining
this identity with (\ref{chevalley}) and the Cartan-Weyl commutation relations. 

The higher-order terms in (\ref{ditqb1n}) are dealt with as in the above proof of (\ref{vneq}).
This proves (\ref{direq}), finishing the proof of Theorem \ref{sqofcoad}.
\enp 

 Finally, we remark that the results in this section have an obvious yet somewhat cumbersome
generalization:
if the orbit $\CO_{\lm}$ is not integral, but such that  $\CO_{\lm/c}$ is integral for some
$c\in\R\backslash \{0\}$, we can construct a strict quantization for the values $\hbar=c/k$,
$k\in\Bbb N$.

\end{document}